\documentclass{elsart}

\usepackage{eepic}

\usepackage[latin1]{inputenc} 
\usepackage{calc}         
\usepackage{graphicx}     
\usepackage{ifthen}       
\usepackage{pst-all}      

\def\leadsto{\hookrightarrow}

\newcommand {\first}[1]{{\it #1}}
\newcommand {\comments} [1]{}

\def\true{{\sl true} }
\def\false{{\sl false}}

\def\sometimes{\mbox{$\exists?$}}
\def\unique{\mbox{$\exists!$}}

\def\existence{{\em existence }}

\def\eu{{\em exists unique }}

\def\es{{\em exists sometimes }}
\def\Es{{\em Exists sometimes }}

\def\DIG{{\cal D}}
\newcommand{\vars}[1]{{\bf V}_{\!#1}}
\newcommand{\lvars}[1]{{\it loc}({#1})}
\newcommand{\type}[1]{{\it type}({#1})}
\def\typeof{{\it type}}

\def\vsigma{\bar{\sigma}}

\def\valpha{{\boldmath\mbox{$\alpha$}}}
\def\vbeta{{\boldmath\mbox{$\beta$}}}

\def\vomega{\bar{\omega}}
\def\vpsi{\bar{\psi}}
\def\vtheta{\bar{\theta}}
\def\veta{\bar{\eta}}

\def\uvec{\bar{u}}

\def\xvec{\bar{x}}

\def\zvec{\bar{z}}
\def\rvec{\bar{r}}
\def\tvec{\bar{z}}
\def\svec{\bar{s}}

\def\yvec{\bar{y}}

\def\yset{{\sf Y}}

\def\lvec{{\sf L}}
\def\wvec{{\sf W}}

\def\uarg{\tilde{u}}

\def\xarg{\tilde{x}}

\def\sarg{\tilde{s}}

\def\xmvec{\bar{x}_{m}}

\def\land{\wedge}
\def\lor{\vee}
\def\implies{\rightarrow}
\def\lequiv{\leftrightarrow}
\def\false{{\sf false}}
\def\true{{\sf true}}

\newcommand {\cnr} [3]{{\small {\noindent
 \begin{tabular}{|l|} \hline #1\\ \hline
 #2\\ \hline #3\\ \hline
          \end{tabular}
}}}

\def\tand{{\sqcap} }

\def\ttop{{\bf 1} }
\def\tbot{{\bf 0} }

\def\minf{{-\infty}}

\def\treal{\mathcal{R}}
\def\tint{\mathcal{Z}}
\def\tlist{{\sf list}}

\newcommand{\iinterval}[2]{{\tint_{[#1,#2]}}}
\newcommand{\rinterval}[2]{{\treal_{[#1,#2]}}}

\def\posint{{\tint_{>0}}}

\def\negint{{\tint_{<0}}}
\def\posreal{{\treal_{>0}}}
\def\poszeroreal{{\treal_{\geq 0}}}

\def\negreal{{\treal_{<0}}}

\def\teq{\equiv}

\def\tleq{\sqsubseteq}

\newtheorem{example}{Example}
\newtheorem{lemma}[example]{Lemma}
\newtheorem{theorem}[example]{Theorem}

\begin{document}

\bibliographystyle{plain}

\begin{frontmatter}

\title{Simplifying Negative Goals Using Typed Existence Properties}
\author{Lunjin Lu}
\address{Department of Computer Science and Engineering, Oakland University}
\ead{lunjin@acm.org}

\author{John G. Cleary}
\address{Department of Computer Science, University of Waikato}
\ead{jcleary@cs.waikato.nz}

\begin{abstract}
A method for extracting positive information from negative goals is
proposed. It makes use of typed existence properties between arguments
of a predicate to rewrite negative goals in a logic program. A typed
existence property is a generalization of functional dependencies in
that an input value maps to a fixed number of output values. Types are
used to specify the domains of the input and output values.  An
implementation of the simplification method is presented and its
complexity is analyzed. A key algorithm of the implementation checks
if an atom in a negative goal can be extracted using a given typed
existence property. A digraph links an atom to the quantified
variables occurring in the atom and is used to quickly retrieve atoms
in the negative goal that may become extractable after some other atom
is extracted.

\end{abstract}

\end{frontmatter}

\section{Introduction} \label{sec:intro}

A challenging issue in logic programming is how to find answers to
negative goals.  Chan introduced the ``constructive negation" rule
which allows non-ground negative goals to bind variables in the same
way as positive ones~\cite{Chan88,Chan89}. Many methods along this
line have been
proposed~\cite{ICLP94:Bossi,ICLP94:Bruscoli,Paola94,Chan88,Chan89,Drabent95,Fages97,Maluszynski89:NACLP,Marchiori96,Moreno-Navarro94,
  Moreno-Navarro96,Moreno-Navarro:2000,Munoz-HernandezMM04,RamirezF93,stuckey95}.
These methods find answers to negative goals by negating a frontier of
a derivation tree for the negated sub-goal.

A different approach was proposed by Cleary that makes use of
\existence properties of arithmetic constraints to rewrite negative
goals~\cite{Cleary97}. There are usually functional dependencies
between arguments to an arithmetic constraint.  Let $add(x,y,z)$
denote $z=x+y$ on the domain of integers, for any integers $x$ and
$y$, then there is a unique $z$ such that $add(x,y,z)$ is true. This
is called an \eu property. It implies that $\neg\exists{z}.add(x,y,z)$
is unsatisfiable and that $\neg\exists z.(add(x,y,z)\land q(z))$ can
be directly simplified to $add(x,y,z)\land\neg q(z)$.  Another kind of
property is called the \es property. Let $log(y,x)$ denote $y=10^x$ on
the domain of integers. Then there is at most one $x$ such that
$log(y,x)$ is true. So, we can directly simplify
$\neg\exists{x}.(log(y,x)\land q(x))$ to $\neg\exists{x}.log(y,x)\lor
log(y,x)\land \neg q(x)$.  The simplification procedure
in~\cite{Cleary97} consists of rewrite rules for these kinds of
property.

     The prerequisite that a functional or partial functional
dependency exists between arguments to a predicate (arithmetic
constraints in~\cite{Cleary97}) is over restrictive. Consider
$sq(x,y)$ in the domain of real numbers where $sq(x,y)$ denotes
$y=x^2$. For every $x$, there is a unique $y$ such that $sq(x,y)$ is
true. However, for every $y>0$, there are two $x$'s such that
$sq(x,y)$ is true. The rewrite rule for \eu properties
in~\cite{Cleary97} doesn't apply directly when it comes to
simplifying $\neg\exists x. (sq(x,y)\land b(x))$.  This
problem is resolved by inserting a tautology $(x\geq 0\lor x<0)$ into
the negative goal and transforming $\neg\exists x. (sq(x,y)\land
b(x))$ into $\neg\exists x_1. (sq(x_1,y)\land x_1\geq 0\land
b(x_1))\land \neg\exists x_2. (sq(x_2,y)\land x_2<0\land b(x_2))$ and
then applying the rewrite rule for \eu properties to the two negative
sub-goals. This causes difficulty because we need to have \eu
properties for complex constraints $(sq(x_1,y)\land x_1\geq 0)$ and
$(sq(x_2,y)\land x_2<0)$. Moreover, inserting a correct tautology, say
$(x\geq 0\lor x<0)$, into the negative goal before rewriting is
rather involved and difficult to mechanise.

This paper generalizes the simplification method in~\cite{Cleary97}
and presents an heuristic implementation of the generalized method.
An input may now correspond to multiple outputs provided that each
output can be isolated into a sub-domain that is expressed as a
type. The generalized method is applicable to more negative goals
because use of types admits more \existence properties and therefore
allows more negative goals to be rewritten. The simplifcation method
can be applied in program transformation because it extracts an atom
from a negative goal without executing the atom. 

A crucial task of any implementation of the generalized method is to
introduce new local (i.e., existentially quantified) variables into an
atom inside a negative goal so that it satisfies a given \existence
property.  Consider $\neg(sq(x,16)\land q(16))$ and this
\existence property.
\begin{quote}
\begin{minipage}[t]{4in}
For any real number $x$, there is a unique real number $y$ such that $sq(x,y)$
is true.
\end{minipage}
\hfill (P)
\end{quote}
 The atom $sq(x,16)$ doesn't satisfy
property (P) and hence cannot be extracted. This is because the
unique $y$ such that $sq(x,y)$ holds is not necessarily $16$.
However, $\neg(sq(x,16)\land q(16))$ can be transformed to
$\neg\exists y'.(sq(x,y')\land 16=y' \land q(16))$ by introducing
a new local variable $y'$. The transformed goal can then be
rewritten to $sq(x,y')\land \neg( 16=y' \land q(16))$ since
$sq(x,y')$ satisfies property (P). In general cases, the task of
introducing new local variables is much more complicated. We
present an algorithm that tests if an \existence property can be
used to extract an atom by introducing zero or more new local
variables.

Another essential issue is how to find quickly an extractable atom
inside a negative goal. Let $G_i$ be $sq(x_{i-1},x_i)$ and $G$ be
$\neg\exists x_1.\exists x_2.\cdots\exists x_n.\exists x_{n+1}.  [G_n
  \land G_{n-1} \cdots G_2\land G_1]$. By repeatedly using property
(P), we can extract from $G$ atoms $G_1$, $G_2$ to $G_{n-1}$ in order
and obtain $G_1 \land G_{2} \cdots G_{n-1}\land \neg\exists
x_{n+1}.G_n$. Observe that $G_j$ becomes extractable after and only
after $x_j$ becomes global upon extraction of $G_{j-1}$. We use a
digraph to represent a negative goal. The digraph links an atom to a
local variable iff the local variable occurs in the atom.  This data
structure allows efficient identification of extractable atoms.

The rest of the paper is organized as follows.  Section~\ref{sec:cn}
presents the generalized simplification method. Section~\ref{sec:digraph}
describes digraphs for representing negative goals and
section~\ref{sec:extractability} presents the algorithm for
introducing new variables. Section~\ref{sec:heuristic} describes briefly the
implementation in ECLipSe Prolog and section~\ref{sec:complexity}
analyzes its complexity.  Section~\ref{sec:related} discusses related
work and section~\ref{sec:conc} concludes. A preliminary version of
this paper appeared in Proceedings of ACM SAC'07, March 11-14, 2007
Seoul, Korea except section~\ref{sec:cn} that is a major revision
of~\cite{ClearyLu98}.

\subsection{Notations}
     We assume that negative goals are of the form $\neg\exists L.G$
     where $L$ is a set of variables and $G$ a conjunction of atoms.
 We also assume that     variables are typed. Expression $y\!\!:\!\!\eta$ indicates that
     variable $y$ has type $\eta$. A type is a finite expression
     denoting a possibly infinite set of terms. We use $\ttop$ to
     denote the set of all ground terms and $\tbot$ the empty set of
     terms. Types $\treal$ and $\tint$ denote the set of real numbers
     and the set of integer numbers respectively. Types $\treal$ and
     $\tint$ with subscripts denote their subtypes. A subscript is
     either an interval or a logical formula.  For instance,
     $\tint_{<0}$ denotes the set of negative integers and
     $\treal_{[0,1)}$ the real interval $[0,1)$.  Relation
         $\sigma\tleq\theta$ holds iff $\sigma$ is a subtype of
         $\theta$; and relation $\sigma\teq\theta$ holds iff $\sigma$
         is equivalent to $\theta$. The intersection of two types
         $\theta$ and $\sigma$ is denoted as $\theta~\tand~\sigma$.
         {We forgo the presentation of a type system because any type
           system for logic programs such
           as~\cite{FruhwirthSVY:LICS91,Mycroft:OKeefe:84,Yardeni:Shapiro:91}
           can be used. We also assume that a set of typed existence
           properties are given.}

      Both existence properties and rewrite rules partition the
      argument list of an atom into several vectors. For an example, let
      add(x,y,z) denote $x+y=z$ where $x,y$ and $z$ range over the
      domain of real numbers.  For given $x$ and $y$, there is exactly
      one $z$ such that add(x,y,z) holds. The input vector $\pi_i$
      consists of the first two arguments $x$ and $y$ and the output
      vector $\pi_o$ consists of the third argument $z$. Formally, a
      vector is a partial function whose domain is a set of argument
      positions (positive integers). Thus, $\pi_i=\{1\mapsto x,
      2\mapsto y\}$ and $\pi_o=\{3\mapsto z\}$. The domain of a vector
      $\pi$ is denoted $dom(\pi)$. The projection of $\pi$ onto
      $D\subseteq dom(\pi)$ is denoted $\pi\downarrow D$.  Then
      $(\pi\downarrow D)(i)=\pi(i)$ if $i\in D$. Otherwise, $(\pi\downarrow D)(i)$ is undefined. We call
      $\pi\downarrow D$ a sub-vector of $\pi$ and accordingly $\pi$
      is a super-vector of $\pi\downarrow D$.  The empty vector is denoted by
      $\epsilon$.  We have $\pi\downarrow \emptyset=\epsilon$ for any
      vector $\pi$.  By an element of a vector $\pi$, we mean $\pi(i)$
      for some $i\in dom(\pi)$.  We use ${\it diff}(\pi)$ to indicate
      that elements in $\pi$ are pair wise different, i.e., ${\it
      diff}(\pi)$ is true iff $\pi(p_1)\neq\pi(p_2)$ for any $p_1\in
      dom(\pi)$ and any $p_2\in dom(\pi)$ such that $p_1\neq p_2$.  In
      the sequel, a letter  with an over bar $\uvec$ denotes a vector of
      different variables, a letter with a tilde $\uarg$
      denotes a vector of terms and a Greek letter with an over bar
      $\veta$ denotes  a vector of types. A vector of types is also called a
      type.  When there is no ambiguity from the context, $\uvec$ is
      also used to denote the set of variables occurring in
      $\uvec$. For instance, put $\xvec=\{1\mapsto x_1,2\mapsto x_2\}$, we write $\exists \xvec. p(\xvec)$
      instead of $\exists x_1.\exists x_2.p(\xvec)$.
      By juxtaposition $\pi_1\pi_2$, we mean that $\pi_1$ and $\pi_2$
      have disjoint domains and $\pi_1\pi_2= \pi_1\cup \pi_2$. For
      instance, $\pi_i\pi_o=\pi_o\pi_i=\{1\mapsto x, 2\mapsto y, 3\mapsto
      z\}$. Let $p$ be of arity $n$. By $p(\pi)$, we mean that
      $dom(\pi)=\{1..n\}$ and $p(\pi)=p(\pi(1),\cdots,\pi(n))$.
      For instance, $add(\pi_i\pi_o)$ stands for $add(x,y,z)$.
      When it is clear from context, a vector is simplify written
      as a sequence with positions omitted.

      By $\uvec:\vsigma$, we mean that $dom(\uvec)=dom(\vsigma)$ and
      $\uvec(i)\!\!:\!\!\vsigma(i)$ for all $i\in dom(\uvec)$.  By
      $\vsigma\tleq\veta$, we mean that $dom(\veta)=dom(\vsigma)$ and
      $\vsigma(i)\tleq\veta(i)$ for all $i\in dom(\vsigma)$. We say
      that $\vsigma$ and $\veta$ intersect iff
      $\vsigma(i)\tand\veta(i)\not\teq\tbot$ for all $ i\in
      dom(\vsigma)$. Let $E$ be an expression. We use $\vars{E}$ to
      denote the set of variables in $E$ and $\type{E}$ the type of
      $E$.

\section{Generalized Method} \label{sec:cn}
This section generalizes the simplification method in~\cite{Cleary97}.  We
first generalize the notion of an \existence property and then the
rewrite rules that make use of \existence properties.

{One rewrite rule applies when it is known that for every
input value a predicate holds for exactly one output value.
Another applies when it is known that for every input value a
predicate holds for at most one output value. It is not necessary
to have the output value available in order to apply these two
rewrite rules. What these two rewrite rules make use of is
knowledge of whether for every input value a predicate holds for
exactly one output value or for at most one output value.}


\subsection{Typed Existence Properties}

An \eu property in~\cite{Cleary97} expresses that, for every
$\uvec$, there is exactly one $\xvec$ such that ${p}(\uvec\xvec)$
holds. In other words, predicate ``p" is be a function from the
domain of $\uvec$ to that of $\xvec$. Parameters in $\uvec$ and
$\xvec$ can be viewed respectively as input and output parameters.
The predicate ``p" may satisfy more than one \eu properties with
different groups of input and output parameters.

As mentioned in section~\ref{sec:intro}, functional dependency is
a strong requirement of a predicate in that many interesting
properties cannot be expressed as functional dependencies. For an
instance, let $exp(x,y)$ denote $y=10^x$ over the domain of real
numbers. Then $exp$ is not a total function from $y$ to $x$ since
there is no $x$ such that $exp(x,y)$ holds for any $y < 0$. This
problem can be resolved by restricting an input to a sub-domain of
its domain. For instance,the property that for every $y>0$ there
is exactly one $x$ such that $exp(x,y)$ holds can be expressed as
$\forall y:\treal_{>0}.\unique x:\treal. exp(x,y)$ where $\unique$
means ``there is exactly one". Types also admits more precise
properties. For instance, the property that for any real number
$x$ there is exactly one non-negative real number $y$ such that
$exp(x,y)$ holds can be expressed as $\forall x:\treal.\unique
y:\treal_{\geq 0}. exp(x,y)$. Another way to generalize the notion
of an \eu property is to allow an input value to correspond to
more than one output value. A typed \eu property of a predicate
thus expresses that for every input value of a given sub-domain,
the predicate holds for a fixed number of output values each of
which can be isolated into a sub-domain. For instance, each
positive number has two square roots one of which is positive and
the other is negative. Formally, a typed \eu property has the
following form where $I$ is a finite set of indices.

\begin{eqnarray} \label{eq:tcomplete}
 \forall\uvec\!\!:\!\!\vsigma.\forall\xvec.[{p}(\uvec\xvec)\implies
      \lor_{i\in{I}} \xvec\in\vtheta_{i}]&&\\
 \label{eq:tunique}
   \forall\uvec\!\!:\!\!\vsigma. \land_{i\in{I}}
   \unique\xvec_{i}\!\!:\!\!\vtheta_{i}.{p}(\uvec\xvec_{i})
\end{eqnarray}
Each $\vtheta_{i}$ is called an \first{output subtype} of the
output parameter $\xvec$. Note that the type of an input parameter
expresses the condition under which a specific property holds.

\begin{example}\label{ex:eu:type}
The fact that, in the domain of real numbers, a positive number has exactly
one negative square root and exactly one positive square root can be
expressed as the following \eu property.
\begin{eqnarray*}
   \forall y\!\!:\!\!\posreal.\forall x.
   (sq(x,y)\implies x\in\posreal\lor x\in\negreal)&&\\
   \forall y\!\!:\!\!\posreal.(\unique x_1\!\!:\!\!\posreal. sq(x_1,y)
    \land \unique x_2\!\!:\!\!\negreal. sq(x_2,y)) &&
\end{eqnarray*}
\end{example}

\begin{example} \label{ex:eu:rq7}
 The fact that the square of any real number is a positive real number is
 expressed as follows.
\begin{eqnarray*}
 \forall x\!\!:\!\!\treal.\forall y.(sq(x,y)\implies y\in\poszeroreal) &&\\
 \forall x\!\!:\!\!\treal.\unique y\!\!:\!\!\poszeroreal. sq(x,y) &&
\end{eqnarray*}
 Note that we have restricted the domain of $y$ to $\poszeroreal$
 rather than $\treal$, which helps avoid the introduction of local
 variables in some cases as explained later.
\end{example}

An \es properties is generalized in the same way, so that every
input value has at most one output value in each of a fixed number
of sub-domains. Formally, a typed \es property is expressed by
(\ref{eq:tcomplete}) and
\begin{equation} \label{eq:tsometimes}
   \forall\uvec\!\!:\!\!\vsigma. \land_{i\in{I}}
   \sometimes\xvec_{i}\!\!:\!\!\vtheta_{i}.{p}(\uvec\xvec_{i})
\end{equation} where  $\sometimes$ denotes ``there is at most one".
Formula (\ref{eq:tsometimes}) requires that, for each $\uvec$ of type $\vsigma$, there
is at most one $\xvec$ in each $\vtheta_i$ such that ${p}(\uvec\xvec)$
holds.  An example of typed \es properties can be found in
Ex.~\ref{ex:svt}.

{
A typed {\em exists} property
\(\forall\uvec\!\!:\!\!\vsigma.\exists\xvec\!\!:\!\!\vtheta.{p}(\uvec\xvec)\)
states that for every $\uvec$ of type $\vsigma$ there are some
$\xvec$ of type $\vtheta$ such that ${p}(\uvec\xvec)$ holds. For
instance, the $append/3$ program satisfies \(\forall
z\!\!:\!\!\tlist(\beta).\exists x\!\!:\!\!\tlist(\beta).\exists
y\!\!:\!\!\tlist(\beta).append(x,y,z)\) which states that every
list $z$ can be split into two lists $x$ and $y$.

A typed miscellaneous property
$\forall\uvec\!\!:\!\!\vsigma.(\neg{p}(\uvec)\lequiv{q}(\uvec))$ states that, for
every $\uvec$ of type $\vsigma$, $\neg{p}(\uvec)$ can be replaced by
${q}(\uvec)$. For instance, we have $\forall
x\!\!:\!\!\tint.y\!\!:\!\!\tint.(\neg(x<y)\lequiv (x\geq y))$.
}

\subsection{Rewrite Rule for Exists Unique Properties}

We now derive a rewrite rule that make uses of typed \existence
properties. Consider first typed \eu properties.  From
(\ref{eq:tcomplete}), we have \( {p}(\uarg\xvec) \lequiv
({p}(\uarg\xvec)\land[\lor_{i\in{I}}\xvec\in\vtheta_i]) \).  Hence
\( ({p}(\uarg\xvec)\land{Q}) \lequiv
(\lor_{i\in{I}}{p}(\uarg\xvec)\land (\xvec\in\vtheta_i)\land{Q}
)\). Distributing $\exists$ over $\lor$, renaming local variables
within their scopes and applying De Morgan's law, we
obtain
\[\neg\exists\xvec\yvec.[{p}(\uarg\xvec)\land{Q}] \lequiv
\land_{i\in{I}} \neg\exists\xvec_{i}\yvec.[{p}(\uarg\xvec_{i})
         \land (\xvec_{i}\in\vtheta_i)\land{Q}[\xvec/\xvec_{i}]]
\] provided that $\vars{\uarg}\cap(\xvec\cup\yvec)=\emptyset$ holds
where ${Q}[\xvec/\xvec_{i}]$ is the result of substituting $\xvec_{i}$ for
$\xvec$ in ${Q}$. Note that {$\xvec$ is renamed into $\xvec_{i}$ for each
output subtype $\vtheta_{i}$.}

The condition $\vars{\uarg}\cap(\xvec\cup\yvec)=\emptyset$ ensures
that $\uarg$ does not contain local variables. To see why this is
necessary, assume the \eu property for integer addition in the
introduction, $\neg\exists y:\tint.(add(x,y,y)\land q(y))$ cannot be
simplified to $add(x,y,y)\land\neg q(y)$ because
$\neg\exists{y:\tint}.add(x,y,y)$ holds for $x\neq 0$. The fact that the
second argument $y$ to $add$ is a local variable invalidates the
condition.

For  $p(\uarg\xvec)$ to be extracted, its
output arguments must satisfy this requirement.
\begin{quote} 
An output argument is a local variable; and for each output subtype
$T_p$ of its corresponding output parameter, either $T_p$ is a subtype
of $T_a$ or $T_p$ doesn't intersect with $T_a$ where $T_a$ is the type
of the output argument.
\end{quote}

\begin{example}
This is an \eu property in the domain of integers.
\begin{eqnarray*}
 \forall x\!\!:\!\!\tint.\forall y\!\!:\!\!\tint.\forall z.(add(x,y,z)\implies z\in\tint) &&\\
 \forall x\!\!:\!\!\tint.\forall y\!\!:\!\!\tint.\unique z\!\!:\!\!\tint. add(x,y,z)&&
\end{eqnarray*}
It states that, for any integers $x$ and $y$, there is a unique integer $z$
such that $add(x,y,z)$ is true. It would be wrong to use the property
to rewrite \( \neg~\exists
z\!\!:\!\!\iinterval{\minf}{10}.(add(10,y\!\!:\!\!\tint,z)\land b(z)) \)
into \( add(10,y\!\!:\!\!\tint,z\!\!:\!\!\iinterval{\minf}{10})\land
\neg~b(z) \).  This is because $z$ can take any value in $\tint$ and
$\iinterval{\minf}{10}$ is not a supertype of $\tint$.
\end{example}

The number of solutions to be negated is limited by the number of
output subtypes of the output parameter. Some output subtypes are
not relevant for a particular negative goal. An output subtype is
relevant iff it intersects with the type of the local variable in
the negative goal. We call an index a \first{relevant index} if
its corresponding output subtype is relevant. We only need to
consider relevant output subtypes when rewriting the negative
goal.
\begin{example}
Let $G$ be
\( \neg\exists x\!\!:\!\!\poszeroreal.(sq(x,y\!\!:\!\!\posreal)\land b(x))
\). From Ex.~\ref{ex:eu:type}, $sq(x,y\!\!:\!\!\posreal)$ has two solutions for $x$,
one of them is in $\negreal$ and the other is in $\posreal$. This
suggests that there are two solutions to be negated. But, the type
$\poszeroreal$ of the local variable $x$ doesn't intersect with
$\negreal$, i.e., only output subtype $\posreal$ is relevant
for $G$. $G$ is rewritten to
\( sq(x_1\!\!:\!\!\posreal,y\!\!:\!\!\posreal)\land \neg~b(x_1)
\)
since the type $\poszeroreal$ of $x$ is a supertype of the
relevant output subtype $\posreal$.
\end{example}

{
The following rewrite rule makes uses of typed {\em exists}
properties. It verifies that an input argument is of the type of the
corresponding input parameter and that the type of an output argument is a
supertype of the type of the corresponding output parameter.

\cnr{ET}{Given $\forall\uvec\!\!:\!\!\vsigma.\exists\xvec\!\!:\!\!\vtheta.{p}(\uvec\xvec)$ and $\type{\uarg}\tleq\vsigma\land
\vars{\uarg}\cap\xvec=\emptyset
\land\vtheta\tleq\veta$}{$\neg\exists\xvec\!\!:\!\!\veta.{p}(\uarg\xvec)\lequiv\false$}

The following miscellaneous rewrite rule verifies that an input argument is
of the type of the corresponding input parameter.

  \cnr{RT}{Given
$\forall\uvec\!\!:\!\!\vsigma.(\neg{p}(\uvec)\lequiv{q}(\uvec))$ and
$\type{\uarg}\tleq\vsigma$}{$\neg{p}(\uarg)\lequiv{q}(\uarg)$}
}

When the requirement on output arguments of an atom is not met, new
local variables need be introduced so that the atom can be extracted.
Consider how an \eu property can be used to rewrite negative goals of
the form
\[ \neg\exists\lvec.  [{p}(\uarg\xarg)\land{Q}]~~~~~~~~~~~~~~~~~~~~~~~~~~~~~~~~~~~~~~~~~~~~~~~~~~~~~~~~~~~~~~~~(g1) \]
 where $\lvec$ is a set of typed
variables. Assume that $\uarg$ is of type $\vsigma$ (I.e.
$\type{\uarg}\tleq\vsigma$) and that variables in $\lvec$ do not occur
in $\uarg$ (I.e. $\vars{\uarg}\cap\lvec=\emptyset$). Then
$\exists\xvec.({p}(\uarg\xvec)\land (\xvec=\xarg)\land Q)$ is
equivalent to $\vee_{i\in{I}}
\exists\xvec_i.({p}(\uarg\xvec_{i}\!\!:\!\!\vtheta_{i})\land
(\xvec_i=\xarg)\land Q)$ from (\ref{eq:tcomplete}). Goal (g1) is
equivalent to $\neg\exists\lvec.\exists\xvec.
[{p}(\uarg\xvec)\land(\xarg=\xvec)\land{Q}]$ and hence is equivalent
to $\neg\exists\lvec. [\vee_{i\in{I}}
\exists\xvec_i.({p}(\uarg\xvec_{i}\!\!:\!\!\vtheta_{i})\land
(\xvec_i=\xarg)\land Q)]$. Distributing $\exists$ over $\vee$,
applying De Morgan's law and using (\ref{eq:tunique}), we deduce that
goal (g1) is equivalent to
\[\land_{i\in{I}}[p(\uarg\xvec_{i}:\vtheta_{i})\land\neg\exists\lvec.((\xarg=\xvec_{i})\land{Q})]
~~~~~~~~~~~~~~~~~~~~~~~~~~~~~~~~~~~(g2) \] provided that (\ref{eq:tcomplete}),
(\ref{eq:tunique}), $\type{\uarg}\tleq\vsigma$ and
$\vars{\uarg}\cap\lvec=\emptyset$ hold.

\begin{example} \label{ex:added2}
Let the \eu property be that in Ex.~\ref{ex:eu:type} and the
negative goal be
\[\neg\exists
z'\!\!:\!\!\tint,x'\!\!:\!\!\treal_{\geq 20}.
(sq(x',y'\!\!:\!\!\treal_{>10})\land Q(x',z')) ~~~~~~~~~~~~~~~~~~~~~~~~~~~~~ (g1')
\] Goal (g1') is an instance of (g1). We have
$\lvec=\{z'\!\!:\!\!\tint, x'\!\!:\!\!\treal_{\geq 20}\}$, $\uarg=y'$
and $\xarg=x'$.  It holds that $y'\in\posreal$ since
$y'\in\treal_{>10}$ and $(\treal_{>10}\tleq\posreal)$. It also holds
that $\vars{\uarg}\cap\lvec=\{y'\}\cap\{z',x'\}=\emptyset$.
Therefore, (g1') rewrites to
\[ \left(\begin{array}{cl}
&    sq(x_1\!\!:\!\!\posreal,y'\!\!:\!\!\treal_{>10}) \wedge \neg\exists
    z'\!\!:\!\!\tint,x'\!\!:\!\!\treal_{\geq 20}.  (x'\!=\!x_1\!\wedge Q(x',z'))
\\ \wedge~& sq(x_2\!\!:\!\!\negreal,y'\!\!:\!\!\treal_{>10})\! \wedge
    \neg\exists z'\!\!:\!\!\tint,x'\!\!:\!\!\treal_{\geq 20}.  (x'\!=\!x_2\!\wedge
    Q(x',z')) \end{array}\right)~~~~~~~(g2')
\]
\end{example}

If $\type{\xarg}$ doesn't intersects with $\vtheta_{k}$ then
$p(\uarg\xvec_{k}\!\!:\!\!\vtheta_{k})\land\neg\exists\lvec.((\xarg=\xvec_{k})\land{Q})$
can be removed from (g2) because $(\xarg=\xvec_{k})$ is
unsatisfiable and any further instantiation of $\xvec_{k}$ has no
effect on the variables of the original goal.   Let
$\wvec$ be the set of those elements of $\lvec$ that occur in
$\xarg$ and $\yset=\lvec\setminus\wvec$. Then
$\neg\exists\lvec.((\xarg=\xvec_{j})\land{Q})$ is equivalent to
 \[ \neg\exists \wvec_j.(\xarg[\wvec/\wvec_j]=\xvec_{j}) \lor
(\xarg[\wvec/\wvec_j]=\xvec_{j}) \land
\neg\exists\yset.{Q}[\wvec/\wvec_j] ~~~~~~~~~~(g3) \] where
$\wvec_j$ is a renaming of $\wvec$. The disequality constraint
$\neg\exists\wvec_j.(\xarg[\wvec\!/\!\wvec_j]=\xvec_{j})$ can be
dealt with by augmenting Chan's simplification procedure with
types.
\begin{example} Continue with Ex.~\ref{ex:added2}.
We have $\wvec=\{x'\!\!:\!\!\treal_{\geq 20}\}$,
$\yset=\{z'\!\!:\!\!\tint\}$ and $J=\{1\}$. The output subtype
$\negreal$ is not relevant since
$(\type{x'}~\tand~\negreal)\teq\tbot$. The sub-formula
\(\neg\exists z'\!\!:\!\!\tint,x'\!\!:\!\!\treal_{\geq 20}.
(x'=x_1\land Q(x',z'))\) in (g2') can be rewritten to
\( \neg\exists
w_1\!\!:\!\!\treal_{\geq 20}.(w_1=x_1) \lor
(w_1\!\!:\!\!\treal_{\geq 20}=x_1)\land\neg\exists
z'\!\!:\!\!\tint. Q(w_1,z') \).
\end{example}

A new local variable is introduced for each output argument in
(g3). As the cost of simplifying
$\neg\exists\wvec_j.(\xarg[\wvec\!/\!\wvec_j]=\xvec_{j})$
increases with the number of equations it contains, it is
desirable to avoid introducing new local variables whenever
possible.  No new local variable need be introduced for an output
argument $r$ if $r$ is a local variable, its type is a super-type
of all relevant output subtypes and it doesn't appear in any
other output argument.

\begin{example} Continue with Ex.~\ref{ex:added2}. \label{ex:added3}
Variable $x'$ is a local variable. Its type is $\treal_{\geq 20}$.
The only relevant output subtype is $\posreal$. A new local
variable was introduced because $\treal_{\geq 20}$ is not a
super-type of $\posreal$.
\end{example}

{
\begin{example}
The following is an \eu property in the domain of integers.
\begin{eqnarray*}
 \forall x\!\!:\!\!\tint. y\!\!:\!\!\tint.\forall z.(add(x,y,z)\implies z\in\tint) &&\\
 \forall x\!\!:\!\!\tint. y\!\!:\!\!\tint.\unique z\!\!:\!\!\tint. add(x,y,z)&&
\end{eqnarray*}
It states that, for any integers $x$ and $y$, there is a unique
integer $z$ such that $add(x,y,z)$ is true. It would be wrong to
use the \eu property to rewrite \( \neg~\exists
z\!\!:\!\!\iinterval{\minf}{10}.(add(10,y\!\!:\!\!\tint,z)\land
b(z)) \) into \(
add(10,y\!\!:\!\!\tint,z\!\!:\!\!\iinterval{\minf}{10})\land
\neg~b(z) \).  This is because $z$ can take any value in $\tint$
and $\iinterval{\minf}{10}$ is not a super-type of $\tint$.
\end{example}
}

{
\begin{example}
The fact that, in the domain of real numbers, a positive number has exactly
one negative square root and exactly one positive square root can be
expressed as the following \eu property.
\begin{eqnarray*}
   \forall y\!\!:\!\!\posreal.\forall x.
   (sq(x,y)\implies x\in\posreal\lor x\in\negreal)&&\\
   \forall y\!\!:\!\!\posreal.(\unique x_1\!\!:\!\!\posreal. sq(x_1,y)
    \land \unique x_2\!\!:\!\!\negreal. sq(x_2,y)) &&
\end{eqnarray*}
Let the negative
goal to rewrite be the following.
\[ \neg\exists x\!\!:\!\!\poszeroreal.(sq(x,y\!\!:\!\!\posreal)\land b(x))
\] where the type of a variable is associated with its first occurrence.
By the above \eu property, $sq(x,y\!\!:\!\!\posreal)$ has
two solutions for $x$, one of them is in $\negreal$ and the
other is in $\posreal$. This suggests that there are two
solutions to be negated. But, the type $\poszeroreal$ of the
local variable $x$ doesn't intersect with $\negreal$, that
is, only output subtype $\posreal$ is relevant for the
negative goal. The negative goal is rewritten to
\[ sq(x_1\!\!:\!\!\posreal,y\!\!:\!\!\posreal)\land \neg~b(x_1)
\]
since the type $\poszeroreal$ of $x$ is a super-type of the
relevant output subtype $\posreal$.
\end{example}
}

 The above considerations lead to the rewrite rule QVT for \eu
 properties in Fig.~\ref{fig:eu}.  The condition
 $\type{\uarg}\tleq\vsigma\land
 \vars{\uarg}\cap(\xvec\cup\yvec)=\emptyset $ in the rewrite rule
 ensures that an input argument is of the type of its corresponding
 input parameter and it doesn't contain any local variables. QVT generates
 only sub-formulae for relevant output subtypes which are
 collected by
 $J=\{i\in{I}~|~(\veta~\tand~\vtheta_{i})\not\teq\tbot\}$.  Variables
 in $\zvec_{j}\rvec_{j}$ and $\wvec_j$ do not occur in the left hand
 side of the rewrite rule. The vector $\zvec_{j}\rvec_{j}$ is typed
 with $\vtheta_j$ while $\wvec_j$ inherits the type of $\wvec$. The
 vector $\rvec$ consists of different variables; and it is a
 sub-vector of $\xarg$ for which no new local variables need be
 introduced.
\begin{figure}
\cnr{QVT}{Given (\ref{eq:tcomplete}), (\ref{eq:tunique}),
   $\type{\uarg}\tleq\vsigma$ and $\vars{\uarg}\cap\lvec=\emptyset$ hold}
    {
$\neg\exists\lvec.[{p}(\uarg\xarg)\land{Q}] \lequiv$\\
~~~~~$\left(\begin{array}{l}
     let~J=\{i\in{I}~|~\type{\xarg}\tand\vtheta_i\not\teq\tbot\}\\
~~~~~\nu\subseteq\{p\mid p\in dom(\xarg) \land \xarg(p)\in\lvec\land \forall j\in{J}.(\vtheta_j(p)\tleq \type{\xarg(p)})\}\\
~~~~~~~~~~~such~that~{\it diff}(\xarg\downarrow\nu)~holds\\
~~~~~\mu=dom(\xarg)\setminus\nu,~~\rvec=\xarg\downarrow\nu,~~~\sarg = \xarg\downarrow\mu\\
     ~~~~~\wvec=(\lvec\cap\vars{\sarg})\setminus\rvec,~~~\yset=\lvec\setminus\wvec\\
    in\\
\land_{j\in{J}}
    \left(\begin{array}{cl}
    & {p}(\uarg[\tvec_{j}\rvec_{j}]\!\!:\!\!\vtheta_{j})
    \land
    \neg\exists\wvec_j.(\sarg[\rvec/\rvec_{j},\wvec/\wvec_j]=\tvec_{j}) \\
     \lor~~~ &
    {p}(\uarg[\tvec_{j}\rvec_{j}]\!\!:\!\!\vtheta_{j})
    \land (\sarg[\rvec/\rvec_{j},\wvec/\wvec_j]=\tvec_{j}) \land
    \neg\exists\yset.{Q}[\rvec/\rvec_{j},\wvec/\wvec_j]
    \end{array}\!\!\right)
\end{array}\right) $}

\caption{\label{fig:eu}Rewrite rule QVT for \eu properties.}
\end{figure}

\begin{example} Continue with Ex.~\ref{ex:eu:type} and Ex.~\ref{ex:added2}.
                QVT rewrites (g1') directly to 
\[\begin{array}{cl}
&    sq(x_1\!\!:\!\!\posreal,y'\!\!:\!\!\treal_{>10}) \land
    \neg\exists w_1\!\!:\!\!\treal_{\geq 20}.(w_1=x_1)
    \\ \lor~~ &
    sq(x_1\!\!:\!\!\posreal,y'\!\!:\!\!\treal_{>10}) \land
    (w_1\!\!:\!\!\treal_{\geq 20}=x_1)\land\neg\exists
     z'\!\!:\!\!\tint. Q(w_1,z')
   \end{array}
\]
\end{example}

\begin{example} \label{ex:append} The {\it append}/3 program satisfies this \eu property.

  \begin{eqnarray*}
  \forall x\!\!:\!\!\tlist(\beta),y\!\!:\!\!\tlist(\beta).z. (append(x,y,z)\implies z\!\!:\!\!\tlist(\beta)) &&\\
 \forall x\!\!:\!\!\tlist(\beta),y\!\!:\!\!\tlist(\beta)\unique z\!\!:\!\!\tlist(\beta).append(x,y,z)&&
  \end{eqnarray*} 
Goal $\neg\exists z\!\!:\!\!\tlist(\beta).(append(x\!\!:\!\!\tlist(\beta),y\!\!:\!\!\tlist(\beta),z),p(z))$ is rewritten to
$
append(x\!\!:\!\!\tlist(\beta),y\!\!:\!\!\tlist(\beta),z\!\!:\!\!\tlist(\beta)),\neg~p(z)$ by QVT.

\end{example}

{ When QVT is  used as a
simplification rule,  it will prune unsatisfiable goals
without doing a satisfiability test.
\begin{example} \label{ex:simp} We have  $\forall y\!\!:\!\!\ttop.x.(x=s(y)\implies x\!\!:\!\!\ttop)$
and $\forall y\!\!:\!\!\ttop.\unique x\!\!:\!\!\ttop.(x=s(y))$ in the
        domain of Herbrand universe. Consider the following
        program.
\begin{verbatim}
p(y).
r(y) :- x=s(y),q(x).
\end{verbatim}
The goal $p(y\!\!:\!\!\ttop),\neg~r(y)$ is reduced to $p(y),\neg\exists
x\!\!:\!\!\ttop.(x=s(y\!\!:\!\!\ttop),q(x))$ which is then simplified
directly into $x\!\!:\!\!\ttop=s(y),p(y\!\!:\!\!\ttop),\neg~q(x)$ using the
above property. Without using this property, $\neg\exists
x\!\!:\!\!\ttop.(x=s(y\!\!:\!\!\ttop),q(x))$ is simplified to
\[ \forall x\!\!:\!\!\ttop.(x \neq s(y\!\!:\!\!\ttop)) \lor (x\!\!:\!\!\ttop=s(y\!\!:\!\!\ttop),\neg~q(x))
\]
and a satisfiability test is then used to eliminate
$\forall~x\!\!:\!\!\ttop.(x\neq s(y\!\!:\!\!\ttop))$. In that sense, the satisfiability
test is pushed into the simplification procedure by the \eu property.
\end{example}
}

\subsection{Rewrite Rule for Exists Sometimes Properties}
The same considerations as in the case for \eu properties lead to the
rewrite rule SVT for \es properties in Fig.~\ref{fig:es}.

\begin{figure}
\cnr{SVT}{Given (\ref{eq:tcomplete}), (\ref{eq:tsometimes}),
$\type{\uarg}\tleq\vsigma$ and $\vars{\uarg}\cap\lvec=\emptyset$
hold} {$\neg\exists\lvec.
    [{p}(\uarg\xarg)\land{Q}] \lequiv$\\
~~~~~$\left(\begin{array}{l}
     let~J=\{i\in{I}~|~\type{\xarg}~\tand~\vtheta_{i}\not\teq\tbot\}\\
~~~~~\nu\subseteq\{p\mid p\in dom(\xarg) \land \xarg(p)\in\lvec\land \forall j\in{J}.(\vtheta_j(p)\tleq \type{\xarg(p)})\}\\
~~~~~~~~~~~such~that~{\it diff}(\xarg\downarrow\nu)~holds \\
~~~~~\mu=dom(\xarg)\setminus\nu,~~~\rvec=\xarg\downarrow\nu,~~~\sarg = \xarg\downarrow\mu\\
     ~~~~~\wvec=(\lvec\cap\vars{\sarg})\setminus\rvec,~~~\yset=\lvec\setminus\wvec\\
    in\\
\land_{j\in{J}} \left(\!\!\begin{array}{ll}
& \neg\exists(\tvec_{j}\rvec_{j})\!\!:\!\!\vtheta_{j}.
{p}(\uarg\tvec_{j}\rvec_{j}) \\ \lor~~&
  {p}(\uarg[\tvec_{j}\rvec_{j}]\!\!:\!\!\vtheta_{j})
   \land \neg\exists\wvec_j.(\sarg[\rvec/\rvec_{j},\wvec/\wvec_j]=\tvec_{j}) \\
 \lor~~&
  {p}(\uarg[\tvec_{j}\rvec_{j}]\!\!:\!\!\vtheta_{j})
                     \land
         (\sarg[\rvec/\rvec_{j},\wvec/\wvec_j]=\tvec_{j}) \land  \neg\exists\yset.{Q}[\rvec/\rvec_{j},\wvec/\wvec_j]
\end{array}\!\!\right)
\end{array}\right)$}
\caption{\label{fig:es}Rewrite rule SVT for \es properties.}
\end{figure}

\begin{example} \label{ex:svt}
The fact that, in the domain of integer numbers, a positive number has at
most one negative square root and at most one positive square root can be
expressed as the following typed \es property.

\begin{eqnarray*}
&& \forall y\!\!:\!\!\posint.\forall x.
   (sq(x,y)\implies x\in\negint\lor x\in\posint)\\
&& \forall y\!\!:\!\!\posint.(\sometimes x_1\!\!:\!\!\negint. sq(x_1,y)\land
            \sometimes x_2\!\!:\!\!\posint. sq(x_2,y) )
\end{eqnarray*}
The local variable $x$ in the negative goal
$\neg\exists{x}\!\!:\!\!\iinterval{0}{20}.(sq(x,y\!\!:\!\!\posint)\land
b(x))$ has a type $\iinterval{0}{20}$ which is not a super-type of
the sole relevant output subtype $\posint$ of the corresponding
output parameter. Therefore, a new local variable $z_2$ of type
$\posint$ is introduced and the negative goal is rewritten to the
following.
\begin{eqnarray*}
&&
\neg\exists z_2\!\!:\!\!\posint. sq(z_2,y\!\!:\!\!\posint) \\
&\lor &
sq(z_2\!\!:\!\!\posint,y\!\!:\!\!\posint)\land~\neg\exists x\!\!:\!\!\iinterval{0}{20}.(x=z_2) \\
&\lor&
sq(z_2\!\!:\!\!\posint,y\!\!:\!\!\posint) \land  (x\!\!:\!\!\iinterval{0}{20}=z_2)\land \neg~b(z_2)
\end{eqnarray*}
\end{example}

Chan's simplification rule can be formalized by a set of \es
properties as follows.
\begin{eqnarray*}
\forall x\!\!:\!\!\ttop.y_1\!\!:\!\!\ttop\cdots y_n\!\!:\!\!\ttop. (x=s(y_1,\cdots,y_n)\implies y_1\in\ttop\land\cdots\land y_n\in\ttop) &&\\
\forall x\!\!:\!\!\ttop. \sometimes y_1\!\!:\!\!\ttop\cdots y_n\!\!:\!\!\ttop.
   (x=s(y_1,\cdots,y_n)) &&
\end{eqnarray*}
These satisfy (\ref{eq:tcomplete}) and (\ref{eq:tsometimes}) and allow
SVT to be applied.

{
There is no rewrite rule with introduction of local variables for {\em
exists} properties because introducing local variables won't lead to
simplification. {Let $\neg\exists\wvec\rvec.{p}(\uarg\sarg\rvec)$ be
the negative goal. Suppose we have
$\forall\uvec\!\!:\!\!\vsigma.\exists\svec\!\!:\!\!\vpsi\rvec\!\!:\!\!\vomega.{p}(\uvec\svec\rvec)$
and $(\type{\uarg}\tleq\vsigma) \land
\vars{\uarg}\cap(\wvec\cup\rvec)=\emptyset\land \vomega\tleq \type{\rvec}$. By
introducing local variables $\zvec\!\!:\!\!\vpsi$, the negative goal
is equivalent to
$\neg\exists\wvec\zvec\rvec.({p}(\uarg\zvec\rvec)\land \zvec=\sarg)$.
Applying (SVT) rewrite rule with the property that $(\zvec=\sarg)$ has
at most one solution, we end
up with \( \neg\exists\wvec\zvec.(\zvec=\sarg)
\lor
(\zvec=\sarg)\land\neg\exists\rvec.{p}(\uarg\zvec\rvec)
\).  The negative goal
$\neg\exists\rvec.{p}(\uarg\zvec\rvec)$
can't be rewritten using (ET) because $\zvec$ are not local variables
in it. Thus, introducing new local variables doesn't help.}
Introduction of local variables is irrelevant to the miscellaneous
rewrite rule as miscellaneous properties have no output parameters.
}

\section{Digraph} \label{sec:digraph}

{
The rewrite rules (ET) and (RT) are applied to negative goals that are
negation of single atom and do not involve introduction of local
variables.  Their implementation is much easier than the other two
rewrite rules and will not be considered.
}

The rewrite rules QVT and SVT can be applied repeatedly to extract
positive information from a negative goal
$\neg\exists\wvec.G_n,\cdots,G_2,G_1$.  A naive implementation would
repeatedly scan a conjunction of goals and check if an atom is
extractable.  After an atom is extracted, some local variables become
global, making it necessary to check if other atoms are
extractable. That would result in an inefficient implementation
because most of those checks would fail.

A previously inextricable atom becomes extractable only after some
of its local variables become global or some of its global
variables are given a value or a smaller type. However, neither
QVT nor SVT changes the type of global variables, nor
will it assign any value to them.  So, after an atom is extracted,
it is only necessary to check those other atoms that share with
the extracted atom some variables that have become global. For
that reason, we use a list $\Phi$ consisting of atoms to be
checked and a digraph $\DIG$ which links each atom with the local
variables it contains. The method repeatedly removes one atom from
$\Phi$ and checks for its extractability until $\Phi$ becomes
empty.  Digraph $\DIG$ is used in order to quickly retrieve the
local variables an atom contains and the atoms containing a
particular local variable. After an atom is extracted, it is moved
out of the scope of the negation and the local variables it
contains become global. This is done by removing the atom and the
local variables from $\DIG$. Before the removal of the local
variables, other atoms linked to them are added to $\Phi$ as their
extractability need to be checked for again.  Initially, every
atom need to be checked.

Let us first consider the case where an \existence property has one
output subtype for its output parameter.  When an atom is extracted by
QVT or SVT without introducing any new local variable, it is moved
out of the scope of the negation and the local variables in it are
promoted to being global. The atom is deleted from $\Phi$ and $\DIG$.
The other atoms that are linked to the local variables are then added
into $\Phi$ and the local variables are deleted from $\DIG$. The
method continues with the updated $\DIG$ and $\Phi$.

\begin{example} \label{ex:digraph}   Let $p$ be
of arity $2$ with the following \eu property.
\[ \forall x\!\!:\!\!\ttop.
    \forall y.(p(x,y)\implies y\in\ttop)~\hspace{1pc}~and~\hspace{1pc}~
 \forall x\!\!:\!\!\ttop.\unique y\!\!:\!\!\ttop.p(x,y)
\]
Let $G_i=p(x_i,x_{i+1})$. The negative goal \(\neg\exists
x_2\!\!:\!\!\ttop.\cdots x_{n+1}\!\!:\!\!\ttop.(G_n, \cdots, G_i, \cdots,
G_1)\) is such that extracting $G_i$ makes $G_{i+1}$ extractable. A naive
implementation of QVT does $\frac{n(n-1)}{2}$ tests by testing
$G_n$ for $n$ times, $G_{n-1}$ for $n-1$ times and so on. The negative goal
has the following graph.

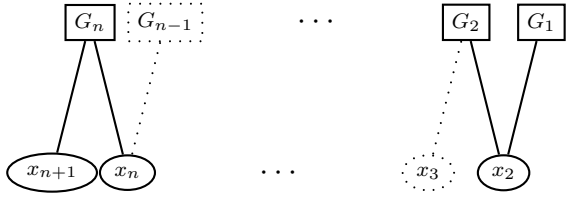
\begin{pspicture}(-2,-.5)(7,3) 
  \rput(.0,.5){\ovalnode{A}{$\scriptstyle x_{n+1}$}}
  \rput(1.0,.5){\ovalnode{B}{$\scriptstyle x_{n}$}}
  \rput(6,.5){\ovalnode{D}{$\scriptstyle x_2$}}

  \rput(.5,2.5){\rnode{E}{\psframebox{$\scriptstyle G_n$}}}
  \rput(5.5,2.5){\rnode{G}{\psframebox{$\scriptstyle G_{2}$}}}
  \rput(6.5,2.5){\rnode{H}{\psframebox{$\scriptstyle G_{1}$}}}

  \ncline{A}{E}
  \ncline{B}{E}
  \ncline{D}{G}
  \ncline{D}{H}

  \psset{linestyle=dotted}

  \rput(1.5,2.5){\rnode{F}{\psframebox{$\scriptstyle G_{n-1}$}}}
  \rput(3.0,.5){$\cdots$} 
  \rput(3.5,2.5){$\cdots$} 
  \rput(5,.5){\ovalnode{C}{$\scriptstyle x_3$}}
  \ncline{B}{F}
  \ncline{C}{G}
\end{pspicture}

The proposed implementation works as follows. Initially, $\Phi$
contains $G_n,\cdots,G_1$ that are removed from
$\Phi$ and tested in that order until $G_1$ is extracted. At that
point, only $G_2$ is added to $\Phi$, it is then immediately removed
and tested. Extracting $G_2$ adds $G_3$ into $\Phi$. This process
continues until $G_n$ is tested and extracted, proving the falsity of
the original negative goal.  A total of $(2n-1)$ tests are performed with
 $G_1$ being tested once and each $G_i$ for $2\leq{i}\leq{n}$ twice.
\end{example}

When an atom is extracted by QVT or SVT by means of introducing
local variables, only some local variables become global and the
derived goals are more complex. However, the residual negative
subgoals can be obtained in the same way as above.

When the output parameter of an \existence property has more than one
output subtype, several complex goals may be derived from the negative
goal. Each of these complex goals may contain a number of residual
negative subgoals to which QVT or SVT may be applicable. However,
these residual negative subgoals differ only in the names and types of
newly promoted global variables. So, the digraph and the checklist for
each of these residual negative subgoals are obtained in the same way.

\section{Extractability}\label{sec:extractability}

Given an atom inside a negation and an \existence property, QVT and
SVT have to decide if the atom satisfies the \existence property
and, if so, decide for which output arguments new local variables need
be introduced.  The rules QVT and SVT differ only in that SVT
has an extra disjunct
$\neg\exists(\tvec_{j}\rvec_{j})\!\!:\!\!\vtheta_{j}.p(\uarg\tvec_{j}\rvec_{j})$
for each relevant output subtype. Otherwise, they are the same.  The
common functionality of QVT and SVT is factored out to a function
$sqvt$. It tests if an atom satisfies an \existence property,
introduces new local variables, decides if an output subtype is
relevant, and renames and types local variables.

An \eu property is represented as follows. Each input parameter
$u\!\!:\!\!\sigma$ in $\uvec\!\!:\!\!\vsigma$ is represented by ${\bf
i}(\sigma)$.  Each output parameter $x$ in $\xvec$ with output subtypes
$\{\theta_{k}~|~k\in{I}\}$ is represented by ${\bf o}(\bar{\Theta})$ where
$\bar{\Theta}$ is a mapping which maps $k$ in ${I}$ to $\theta_{k}$.  An
\eu property has the following representation where input and output
parameters may be interspersed.
\[ \langle{p}(\cdots,{\bf i}(\sigma),\cdots,{\bf o}(\bar{\Theta}),\cdots),I\rangle
\]
The set of \eu properties is denoted by $\Gamma_{!}$. We use the same
representation for an \es property and denote the set of \es properties by
$\Gamma_{?}$.

\begin{example} \label{ex:rep}
The \es property in Ex.~\ref{ex:svt} is represented by this item in $\Gamma_{?}$:
\( \langle
sq({\bf o}(\{1\mapsto\negint,2\mapsto\posint\}),{\bf i}(\posint)),
\{1,2\}\rangle \).
\end{example}

\begin{figure}[t]

\vbox{\noindent
function $sqvt(P,G,\lvec)$

begin
\begin{itemize}
\item [(01)] Let $G$ be $q(\cdots,t_{u},\cdots, t_{x},\cdots)$ and
      $P$ be $\langle{p}(\cdots,{\bf i}(\sigma),\cdots, {\bf o}(\bar{\Theta}),\cdots),I\rangle$;
\item [(02)] if $q=p$ and $(\type{t_{u}}\tleq\sigma)\land
(\vars{t_{u}}\cap\lvec)=\emptyset$
      for each $t_{u}$ matching an ${\bf i}(\sigma)$
\item [(03)]  then
       \begin{itemize}
        \item [(04)] \(\rvec:=\epsilon;\xmvec:=\epsilon;\tvec:=\epsilon;\sarg:=\epsilon;\wvec:=nil;J:=I\);
    \item[(05)]
 for each $t_{x}$ matching an ${\bf o}(\bar{\Theta})$
 do
  $J:=J\cap\{k \mid (\type{t_{x}}~\tand~\bar{\Theta}(k))\not\teq\tbot\}$ od;

    \item [(06)]
       for each $t_{x}$ at position p matching an ${\bf o}(\bar{\Theta})$ do
           \begin{itemize}
         \item [(07)]
           if $t_{x}\in(\lvec\setminus\rvec)\land
              \forall j\in J.(\bar{\Theta}(j)\tleq \type{t_{x}})$
           \item [(08)]   then
            \begin{itemize}
            \item [(09)] $\rvec:=\rvec[p\mapsto t_{x}]$;
            \item [(10)] $\xmvec:=\xmvec[p\mapsto (t_{x},\bar{\Theta})]$;
            \end{itemize}
        \item [(11)]   else
            \begin{itemize}
            \item [(12)] $\sarg:=\sarg[p\mapsto t_{x}]$;
                \item [(13)] for each
          $v\in((\vars{t_{x}}\cap\lvec)\setminus(\rvec\cup\wvec))$
 do $\wvec:=v::\wvec$ od;
            \item [(14)] $z:=newv(\ttop)$; $\tvec:=\tvec[p\mapsto
            z]$;                $G:=G[t_{x}/z]$;
            \item [(15)] $\xmvec:=\xmvec[p\mapsto (z,\bar{\Theta})]$
            \end{itemize}
             \item  [(16)] fi;
             \end{itemize}
         \item  [(17)] od;
             \item [(18)]  $\xvec:=map(fst,\xmvec)$;
         \item [(19)]  $\xvec_{cs}:=\bigcup_{j\in{J}}\{map(\lambda e. newv((snd(e))(j)),\xmvec)\}$;
          \item [(20)] return $(G,\xvec,\xvec_{cs},\sarg,\tvec,\rvec,\wvec)$
    \end{itemize}
\item [(21)] else return $nil$
\item [(22)] fi
\end{itemize}
end;
}
\caption{\label{fig:sqvt} The $sqvt$ function where $x::L$ is
a list with head $x$ and tail $L$.}
\end{figure}

Fig.~\ref{fig:sqvt} defines $sqvt$ with the following auxiliary
functions. A call to $newv(T)$ creates a new variable of type $T$.
Given a pair, the function ${\it fst}$ returns the first component
while $snd$ returns the second.  The high order function
$map$ applies a function $f$ to a vector $\pi$ point-wise:
$map(f,\pi)(i) = f(\pi(i))$ for each $i\in dom(\pi)$ and
$dom(map(f,\pi))= dom(\pi)$.

Given an \existence property $P$ of the form
$\langle{p}(\cdots,{\bf i}(\sigma),\cdots, {\bf
o}(\bar{\Theta}),\cdots),I\rangle$ and an atom $G$ of the form
$q(\cdots,t_{u},\cdots, t_{x},\cdots)$ and a set $\lvec$ of local
variables, $sqvt$ first checks if it is possible to replace some
output arguments in $G$ with newly introduced local variables so
as to make $G$ satisfy $P$.  { Since a new local variable
can be introduced for any output argument in $G$, $G$ can be made
to satisfy $P$ if $q=p$ and each of its input argument is of the
type specified by $P$ and contains no local variable.} The test is
done in line (02). Function $sqvt$ returns $nil$ from line (21)
if this test fails. Otherwise, $sqvt$ classifies every output
argument according to whether a new local variable need be
introduced for it or not. The variable $\rvec$ holds the vector of
output arguments for which no new local variables need be
introduced, $\sarg$ is the vector of other output arguments and
$\tvec$ is the vector of corresponding newly introduced local
variables. Whenever a new local variable $z$ is introduced for an
output argument $t_{x}$, $sqvt$ substitutes $z$ for $t_{x}$ in
$G$.  The function $sqvt$ collects the list $\wvec$ of the local
variables that occur in $\sarg$ but not in $\rvec$. It also builds
up the vector $\xmvec$ of the new output arguments each of which
is associated with a mapping from indices in ${I}$ to types and
collects the set ${J}$ of relevant indices for $G$. Line (04)
initializes these vectors and sets. Line (05) computes the set
${J}$ of relevant indices. The (06)-(17) loop iterates through all
output arguments.  Line (07) determines if it is necessary to
introduce a new local variable for the output argument $t_{x}$
under consideration.  If not, line (09) adds $t_{x}$ into $\rvec$
and line (10) adds to $\xmvec$ a pair consisting of $t_{x}$ and
the mapping for the corresponding output parameter in $P$.
Otherwise, line (12) adds $t_{x}$ to $\sarg$, line (13) adds to
$\wvec$ the local variables in $t_{x}$ that do not occur in
$\rvec$ or $\wvec$, line (14) introduces a new local variable $z$
of type $\ttop$, adds $z$ to $\tvec$ and substitutes $z$ for
$t_{x}$ in $G$, and line (15) adds to $\xmvec$ a pair consisting
of $z$ and the mapping for the corresponding output parameter in
$P$. The newly introduced local variable $z$ in line (14) will be
renamed and attached with an appropriate type from the mapping
paired with it in $\xmvec$. Line (18) extracts the vector $\xvec$
of the new output arguments of $G$. Line (19) makes, for each
relevant index in ${J}$, a new copy of $\xvec$ and types the copy
with an appropriate type, and collects the set $\xvec_{cs}$ of all
the copies made. For a fixed index $j\in{J}$, line (19) does the
following for each pair in $\xmvec$. It first takes the second
component of the pair which is a mapping from indices to types,
then finds the type for the index $j$, and creates a new variable
of that type. Line (20) returns with required information.

\begin{example} \label{ex:sqvt} Continue with Ex.~\ref{ex:rep}.
Let $G=sq(x\!\!:\!\!\iinterval{0}{20},y\!\!:\!\!\posint)$ and
$\lvec=\{x\!\!:\!\!\iinterval{0}{20}\}$.  Then
$sqvt(P,G,\lvec)=(G',\xvec,\xvec_{cs},\sarg,\tvec,\rvec,\wvec)$
with \(G'=sq(z\!\!:\!\!\ttop,y\!\!:\!\!\posint)\),
\(\xvec=z\!\!:\!\!\ttop\),
\(\xvec_{cs}=\{z_{2}\!\!:\!\!\posint\}\),
$\sarg=x\!\!:\!\!\iinterval{0}{20}$, \(\tvec=z\!\!:\!\!\ttop\),
$\rvec=\epsilon$, and $\wvec=\{x\!\!:\!\!\iinterval{0}{20}\}$.
\end{example}

\begin{lemma}
The time complexity of the test for the extractability of an atom with respect to an \eu or \es property
is linear in the size of the atom.

{\it Proof.} The time complexity of the function $sqvt$ is
proportional to the size of the atom, given an atom and an
\existence property. When an atom is tested for its
extractability, it may be necessary to match it against several
different \existence properties before it can be decided whether
or not it is extractable. When it is not extractable, it has to be
matched against all those \existence properties that have the same
predicate symbol as the atom. The number of the \existence
properties that have the same predicate symbol as an atom is
bounded, which implies the time complexity of the test for the
extractability of an atom is proportional to the size of the atom.

\end{lemma}

The following theorem gives the correctness of $sqvt$. In addition, it
states that $sqvt$ introduces a new variable only when it is
necessary.
\begin{theorem} Let $P$ be an \es (resp. \eu) property, $G$ an atom, $Q$ a conjunction of goals and
$\lvec$ a set of variables.

\begin{itemize}
\item [a)] Atom $G$ can be extracted from $\exists\lvec.(G\land
Q)$ by SVT (resp. QVT) using $P$ iff $sqvt(P,G,\lvec)\neq{nil}$.
\end{itemize}
Furthermore, letting
$sqvt(P,G,\lvec)=(G',\xvec,\xvec_{cs},\sarg,\tvec,\rvec,\wvec)$,
\begin{itemize}
\item [b)] $\rvec$, $\sarg$, $\zvec$ and $\wvec$ are as in SVT
(resp. QVT) and $\rvec$ is maximal in the sense that any proper
super-vector of $\rvec$ will include at least one output argument
of $G$ for which a new variable must be introduced;

 \item [c)]
$G'=G[\sarg/\zvec]$;

 \item [d)] $\xvec$ is the vector of the output
arguments of $G'$; and

 \item [e)] $\xvec_{cs}$ is a set of vectors with
each being a fresh copy of
      $\xvec$ typed by an output subtype of $P$ that is relevant to $G$.
\end{itemize}

{\it Proof.} Postulate (a) follows from the conditional statement
beginning at line (02). Line (05) computes the set $J$ of relevant
indices since two vectors of types  with the same domain intersect
iff their corresponding components at each position in the domain
intersect. The logic of the loop beginning at line (06) ensures
that $\rvec$, $\sarg$ and $\zvec$ are computed correctly without
computing their corresponding sets of indices and it also ensures
the maximality of $\rvec$. Therefore, postulates (b) and (c) hold.
The postulates (d) and (e) follow from lines (18) and (19)
respectively.

\end{theorem}

\section{Implementation} \label{sec:heuristic}


With a negative goal being represented by $neg(\Phi,\DIG)$ where
$\Phi$ is the checklist and $\DIG$ is the digraph,  QVT
and SVT are implemented as a derivation rule $\leadsto_{sqvt}$ {which
derives from the lefthand side of QVT (respectively SVT) each
conjunt in a disjunctive normal form of the righthand side of QVT
(respectively SVT).}
 {Let $\lvars{\DIG}$ be the set of local
  variables in $\DIG$, $delete(Ns,\DIG)$ be the result of deleting
  nodes in $Ns$ from $\DIG$, $link(N,Ns,\DIG)$ be true iff $\DIG$
  links node $N$ with some node in $Ns$.

\begin{itemize}
\item
\( \valpha,neg(\{G\}\cup\Phi,\DIG),\vbeta
\leadsto_{sqvt} \valpha,N_{l},\vbeta
\)  for each $1\leq{l}\leq{k}$ if\\ $\exists{P}\in\Gamma_{!}.sqvt(P,G,\lvars{\DIG})=(G',\xvec,\xvec_{cs},\sarg,\tvec,\rvec,\wvec)$ and $N_{1}\lor N_{2}\lor\cdots\lor N_{k}$ is a disjunctive normal form of
    \[
         \land_{\xvec'\in\xvec_{cs}}
  \left[\begin{array}{l}
    let~\wvec'=map(newv\circ{\typeof},\wvec)~in\\
           \left(\begin{array}{ll}
     &     G' \land
           \neg\exists\wvec'.(\sarg[\wvec/\wvec']=\tvec) )[\xvec/\xvec'] \\ 
\lor &
          G' \land
           (\sarg[\wvec/\wvec']=\tvec))[\xvec/\xvec']~ \land~ neg(\Phi',\DIG')[\wvec/\wvec'][\xvec/\xvec']

         \end{array}\right)
 \end{array}\right] \]
     where $\Phi'=\Phi\cup\{N~|~link(N,\rvec\cup\wvec,\DIG)\}\setminus\{G\}$
     and $\DIG'=delete(\rvec\cup\wvec\cup\{G\},\DIG)$.
    The above formula corresponds to the righthand side of QVT in that
    $\xvec'$ corresponds to $\zvec_j\rvec_j$ and $\wvec'$ to $\wvec_j$.
    Note that $\xvec'$ and $\wvec'$ are typed when they are created.
\item
\( \valpha,neg(\{G\}\cup\Phi,\DIG),\vbeta
\leadsto_{sqvt} \valpha,N_{l},\vbeta
\)  for each $1\leq{l}\leq{k}$ if\\ $\exists{P}\in\Gamma_{?}.sqvt(P,G,\lvars{\DIG})=(G',\xvec,\xvec_{cs},\sarg,\tvec,\rvec,\wvec)$ and $N_{1}\lor N_{2}\lor\cdots\lor N_{k}$ is a disjunctive normal form of
    \[
         \land_{\xvec'\in\xvec_{cs}}
 \left[\begin{array}{l}
    let~\wvec'= map(newv\circ{\typeof},\wvec)~in\\
     \left(\begin{array}{ll}
       &  \neg\exists\xvec'.G'[\xvec/\xvec']\\ \lor &
         G' \land
          \neg\exists\wvec'.(\sarg[\wvec/\wvec']=\tvec))[\xvec/\xvec'] \\ \lor&
         G' \land (\sarg[\wvec/\wvec']=\tvec))[\xvec/\xvec']~\land~ neg(\Phi',\DIG')[\wvec/\wvec'][\xvec/\xvec']
         \end{array}\right)
  \end{array}\right]\]
     where $\Phi'=\Phi\cup\{N~|~link(N,\rvec\cup\wvec,\DIG)\}\setminus\{G\}$
     and $\DIG'=delete(\rvec\cup\wvec\cup\{G\},\DIG)$.
\item \( \valpha,neg(\{G\}\cup\Phi,\DIG),\vbeta
\leadsto_{sqvt} \valpha,neg(\Phi,\DIG),\vbeta\)\\ if $\forall P\in\Gamma_{!}\cup\Gamma_{?}.sqvt(P,G,\lvars{\DIG})=nil$. This rule removes from the checklist an atom which doesn't satisfy any
\existence property.
\item \( \valpha,neg(\emptyset,\Lambda),\vbeta\leadsto_{sqvt}\false \)
where $\Lambda$ is the empty digraph. Note that $neg(\emptyset,\Lambda)$
represents $\neg\true$.
\end{itemize}
}

{

\begin{example}
The goal $\neg\exists
x\!\!:\!\!\rinterval{-20}{20}.u\!\!:\!\!\poszeroreal.(sq(x,y\!\!:\!\!\posreal)\land
add(x,u,-1))$ is represented as $F_0$ below where the checklist is depicted as a group of pointers to atoms.
\[F_0=
neg(
\mbox{\setlength{\unitlength}{0.00058333in}
\begin{picture}(2444,680)(0,100)
\thicklines
\path(687,350)(687,127)
\path(701,140)(1751,365)
\path(1759,367)(1759,142)
\spline(27,143)
(312,675)(1235,675)(1684,503)
\path(1438.418,532.824)(1684.000,503.000)(1481.345,644.883)
\spline(192,158)(312,383)(612,405)
\path(377.031,327.608)(612.000,405.000)(368.255,447.286)
\put(627,375){{\tiny $sq(x\!,\!y\!\!:\!\!{\treal}_{>0})$}}
\put(627,0){{\tiny $x\!\!:\!\!{\treal}_{[-20,20]}$}}
\put(1602,0){{\tiny $u\!\!:\!\!{\treal}_{\tiny\geq 0}$}}
\put(1602,375){{\tiny add(x,u,-1)}}
\end{picture}
}
)
\]

Using (1') and (2'), we have \( F_0\leadsto_{sqvt}F_1\), \(
F_0\leadsto_{sqvt}F_2\), \( F_0\leadsto_{sqvt}F_3\) and \(
F_0\leadsto_{sqvt}F_4\) and $\{F_1,F_2,F_3,F_4\}$ is a frontier of $F_0$
where
\begin{eqnarray*}
 F_1 &=& sq(z_1\!\!:\!\!\negreal,y\!\!:\!\!\posreal)\land\neg\exists x_1\!\!:\!\!\rinterval{-20}{20}.(x_1=z_1)\\
 &&  \land sq(z_2\!\!:\!\!\posreal,y)\land\neg\exists x_2\!\!:\!\!\rinterval{-20}{20}.(x_2=z_2)\\
 F_2 & = & sq(z_1\!\!:\!\!\negreal,y\!\!:\!\!\posreal)\land\neg\exists x_1\!\!:\!\!\rinterval{-20}{20}.(x_1=z_1) \\
 &&    \land sq(z_2\!\!:\!\!\posreal,y)\land (x_2\!\!:\!\!\rinterval{-20}{20}=z_2) \land F_6\\
 F_3 &=& sq(z_1\!\!:\!\!\negreal,y\!\!:\!\!\posreal)\land (x_1\!\!:\!\!\rinterval{-20}{20}=z_1)\land F_5 \\
   &&   \land  sq(z_2\!\!:\!\!\posreal,y)\land\neg\exists x_2\!\!:\!\!\rinterval{-20}{20}.(x_2=z_2) \\
 F_4 &=& sq(z_1\!\!:\!\!\negreal,y\!\!:\!\!\posreal)\land (x_1\!\!:\!\!\rinterval{-20}{20}=z_1)\land F_5 \\
 &&    \land sq(z_2\!\!:\!\!\posreal,y)\land (x_2\!\!:\!\!\rinterval{-20}{20}=z_2) \land F_6
\end{eqnarray*}
with
\begin{eqnarray*}
F_5 &=&
neg(
\mbox{\setlength{\unitlength}{0.00058333in}
\begin{picture}(2444,680)(0,100)
\thicklines
\path(1459,367)(1459,142)
\spline(27,143)(312,675)(1235,675)(1684,503)
\path(1438.418,532.824)(1684.000,503.000)(1481.345,644.883)
\put(1202,0){{\tiny $u\!\!:\!\!{\treal}_{\geq 0}$}}
\put(702,375){{\tiny $add(x_1\!\!:\!\!{\treal}_{[-20,20]},u,-1)$}}
\end{picture}
}
)\\
F_6 &=&
neg(
\mbox{\setlength{\unitlength}{0.00058333in}
\begin{picture}(2444,680)(0,100)
\thicklines
\path(1459,367)(1459,142)
\spline(27,143)(312,675)(1235,675)(1684,503)
\path(1438.418,532.824)(1684.000,503.000)(1481.345,644.883)
\put(1202,0){{\tiny $u\!\!:\!\!{\treal}_{\geq 0}$}}
\put(702,375){{\tiny $add(x_2\!\!:\!\!{\treal}_{[-20,20]},u,-1)$}}
\end{picture}
}
)
\end{eqnarray*}

The following is an \eu property for addition.
\[ \forall x\!\!:\!\!\treal. y\!\!:\!\!\treal.z.(add(x,z,y)\implies z\in\treal)
\]
\[ \forall x\!\!:\!\!\treal. y\!\!:\!\!\treal.\unique z\!\!:\!\!\treal. add(x,z,y)
\]
Using this property, we have
\\ $F_5\leadsto_{sqvt}
add(x_1\!\!:\!\!\rinterval{-20}{20},v_1\!\!:\!\!\treal,-1)\land\neg\exists u_1\!\!:\!\!\poszeroreal.(u_1=v_1)$ and
$F_5\leadsto_{sqvt}add(x_1\!\!:\!\!\rinterval{-20}{20},v_1\!\!:\!\!\treal,-1)\land (u_1\!\!:\!\!\poszeroreal=v_1) \land
neg(\emptyset,\Lambda)$ and  $F_6\leadsto_{sqvt}
add(x_2\!\!:\!\!\rinterval{-20}{20},v_2\!\!:\!\!\treal,-1)\land\neg\exists u_2\!\!:\!\!\poszeroreal.(u_2=v_2)$ and
$F_6\leadsto_{sqvt}add(x_2\!\!:\!\!\rinterval{-20}{20},v_2\!\!:\!\!\treal,-1)\land (u_2\!\!:\!\!\poszeroreal=v_2) \land
neg(\emptyset,\Lambda)$. Since the subgoals $\neg\exists
u_1\!\!:\!\!\poszeroreal.(u_1=v_1\!\!:\!\!\treal)$ and $ \neg\exists u_2\!\!:\!\!\poszeroreal.(u_2=v_2\!\!:\!\!\treal)$ are equivalent
to type constraints $v_1\!\!:\!\!\negreal$ and $v_2\!\!:\!\!\negreal$ respectively and
$neg(\emptyset,\Lambda)$ is unsatisfiable, $\{add(x_1\!\!:\!\!\rinterval{-20}{20},v_1\!\!:\!\!\negreal,-1)\}$ is a
frontier of $F_5$ and $\{add(x_2\!\!:\!\!\rinterval{-20}{20},v_2\!\!:\!\!\negreal,-1)\}$ is a frontier of $F_6$.

$\neg\exists x_1\!\!:\!\!\rinterval{-20}{20}.(x_1=z_1\!\!:\!\!\negreal)$ is
equivalent to type constraint $z_1\!\!:\!\!\mbox{$\treal_{<(-20)}$}$, and
$\neg\exists x_2\!\!:\!\!\rinterval{-20}{20}.(x_2=z_2\!\!:\!\!\posreal)$ to
$z_2\!\!:\!\!\mbox{$\treal_{>20}$}$, Solving
$(x_1\!\!:\!\!\rinterval{-20}{20}=z_1\!\!:\!\!\negreal)$ restricts the
types of both $x_1$ and $z_1$ to \mbox{$\treal_{[-20,0)}$} whilst solving
$(x_2\!\!:\!\!\rinterval{-20}{20}=z_2\!\!:\!\!\posreal)$ restricts the
types of both $x_2$ and $z_2$ to $\mbox{$\treal_{(-20,0]}$}$. Therefore
$\{F_7,F_8,F_9,F_{10}\}$ is a frontier of $F_0$ where
\begin{eqnarray*}
 F_7 &=& sq(z_1\!\!:\!\!\mbox{$\treal_{<(-20)}$},y\!\!:\!\!\posreal) \land sq(z_2\!\!:\!\!\mbox{$\treal_{>20}$},y)\\
 F_8 &=& sq(z_1\!\!:\!\!\mbox{$\treal_{<(-20)}$},y\!\!:\!\!\posreal)
     \land sq(z_2\!\!:\!\!\mbox{$\treal_{(-20,0]}$},y)  \land add(z_2,v_2\!\!:\!\!\negreal,-1)\\
 F_9 &=& sq(z_1\!\!:\!\!\mbox{$\treal_{[-20,0)}$},y\!\!:\!\!\posreal)\land add(z_1,v_1\!\!:\!\!\negreal,-1)
      \land  sq(z_2\!\!:\!\!\mbox{$\treal_{>20}$},y\!\!:\!\!\treal) \\
 F_{10} & =& sq(z_1\!\!:\!\!\mbox{$\treal_{[-20,0)}$},y\!\!:\!\!\posreal)\land add(z_1,v_1\!\!:\!\!\negreal,-1) \\
     && \land sq(z_2\!\!:\!\!\mbox{$\treal_{(-20,0]}$},y)\land add(z_2,v_2\!\!:\!\!\negreal,-1)
\end{eqnarray*}
Note that none of $F_{7}, F_{8}, F_{9}$ and $F_{10}$ contain a negation!
\end{example}
}


We have implemented in ECLiPSe~\cite{Aggoun+95} a prototype
simplification system that also implements Chan's constructive
negation rule. A type is associated with a variable as an
attribute~\cite{Brisset+94}.  The top-level of the simplification
system is neg/2. $neg(G,L)$ is true iff $\neg \exists L.G$ is true. It
constructs a digraph representation for $\neg \exists L.G$ and applies
$\leadsto_{sqvt}$ repeatedly until no rewriting can be done. It then
displays the derived goal.

\begin{example} \label{ex:session}
This example illustrates a session with the prototype.  \comments{$type\_set(Var,Type)$ sets the type of the
variable $Var$ to $Type$.}  Term $real(l,u)$ encodes type
$\treal_{[l,u]}$.

\begin{verbatim}
[eclipse 2]: neg((sq(X:real(-0.5,0.5),U), sq(Y:real(-1,1),V),
                  add(U,V,W:real(0,1))), [U,V]).

sq(Y:real(-1, 1), V1:real), add(Z:real, V1:real, W:real(0, 1)),
sq(X:real(-0.5, 0.5), U1:real), neg_eq(Z:real, U1:real, []);

no (more) solution.
\end{verbatim}

I.e.,
\( \neg \exists
U:\ttop.V:\ttop. (sq(X\!\!:\!\!\treal_{[-0.5,0.5]},U),
sq(Y\!\!:\!\!\treal_{[-1,1]},V), add(U,V,W\!\!:\!\!\treal_{[0,1]}))\)
rewrites to
\( sq(Y\!\!:\!\!\treal_{[-1, 1]}, V1\!\!:\!\!\treal),
add(Z\!\!:\!\!\treal, V1, W\!\!:\!\!\treal_{[0, 1]}),
sq(X\!\!:\!\!\treal_{[-0.5, 0.5]}, U1\!\!:\!\!\treal), Z \neq U1\).
The prototype incorporates  \existence properties of arithmetic
constraints. The programmer may provide
\existence properties as in the following.

{\small
\begin{verbatim}
[eclipse 3]: declare_existence_property(
  eu(append(i(list(Beta)),i(list(Beta)),o([(1,list(Beta))])),[1])),
declare_existence_property(
  eu(sort(i(list(Gamma)),o([(1,list(Gamma))])),[1])),
neg((append(X:list(real),Y:list(real),Z),sort(Z,W), b(W)),[W,Z]).

append(X:list(real),Y:list(real),Z:list(real)),sort(Z,W),neg(b(W),[]). 

no (more) solution.
\end{verbatim}
}
\end{example}

{
\begin{example}
{\small
\begin{verbatim}
[eclipse 1]: RGt0 =  and(real(0,pinf), not(real(0,0))),
             type_set(X,real(-20,20)), type_set(Y, RGt0),
             type_set(U,real(0,pinf)),
             neg([Y],(sq(X,Y),add(X,U,-1))),
             delayed_goals(L), print(L),nl. 
 
sq(Z1:real(0,pinf) and not(real(0,20)),Y:real(0,pinf) and not(real(0,0))), 
sq(Z2:real(minf,0) and not(real(-20,0)), Y:real(0,pinf) and not(real(0,0)));

sq(Z1:real(0,pinf) and not(real(0,20)), Y:real(0,pinf) and not(real(0,0))), 
sq(Z2:real(-20,0) and not(real(0,0)), Y:real(0,pinf) and not(real(0,0))), 
add(Z2:real(-20,0) and not(real(0,0)), V2:real and not(real(0,pinf)),-1);

sq(Z1:real(0,20) and not(real(0,0)), Y:real(0,pinf) and not(real(0,0))), 
add(Z1:real(0,20) and not(real(0,0)), V1:real and not(real(0,pinf)),-1), 
sq(Z2:real(minf,0) and not(real(-20,0)), Y:real(0,pinf) and not(real(0,0)))

sq(Z1:real(0,20) and not(real(0,0)), Y:real(0,pinf) and not(real(0,0))), 
add(Z1:real(0,20) and not(real(0,0)), V1:real and not(real(0,pinf)), -1), 
sq(Z2:real(-20,0) and not(real(0,0)), Y:real(0,pinf) and not(real(0,0))), 
add(Z2:real(-20,0) and not(real(0,0)), V2:real and not(real(0,pinf)), -1)

no (more) solution.
\end{verbatim}
}
\end{example}
}

\section{Time Complexities} \label{sec:complexity}

Given a negative goal, a $\leadsto_{sqvt}$ derivation step extracts an
atom out of a negation and produces several residual negative goals
which are then processed in subsequent derivation steps. The time
complexity of $\leadsto_{sqvt}$ with respect to a negative goal is
measured by the time spent on all possible derivations from the
negative goal.

Our analysis is based on a notion of a spawning tree $SPT_G$ for a
negative goal $G$. The nodes in $SPT_G$ are negative goals that are
derived from $G$ by repeated applications of $\leadsto_{sqvt}$. Let
$G'$ be a node $SPT_G$ and $G''$ occurs in one of the conjunctive
goals derived from $G'$ by $\leadsto_{sqvt}$. Then $G''$ is a child of
$G'$.

Let the negative goal $G$ consist of $m$ atoms with non-decreasing
sizes $s_{i},1\leq i\leq m$.  Consider the time complexity of
$\leadsto_{sqvt}$. We weight the $i^{th}$ atom in $G$ by the number
$w_{i}$ of those atoms that share local variables with the $i^{th}$
atom and are smaller in size than the $i^{th}$ atom.

Some branches in $SPT_G$ result from failed extractability tests. The
parent node linked by such a branch has exactly one child and is
called futile. Other nodes correspond to successful extractability
tests and are called fruitful. The set of fruitful nodes in $SPT_G$ is
dentoed $Fr(SPT_G)$.  Let $s_{nd}$ is the size of the atom that is
extracted at a fruitful node $nd$ and $w_{nd}$ the weight of the atom.

\begin{theorem} Let $G$ be a negative goal.
\begin{enumerate}
\item 
The time cost of the extractability tests performed along a path
in $SPT_G$ is $\mathcal{O}(\Sigma_{i=1}^m (w_{i}+1)\times s_i)$.
\item 
 The time cost
 of all $\leadsto_{sqvt}$ derivations from $G$ is
 \(\mathcal{O}(\Sigma_{nd\in Fr(SPT_G)} (w_{nd}+1)*s_{nd})\).
\end{enumerate}

{\it Proof.} {Consider (1) first. We only need to consider the worst case where
          each atom in $G$ will finally be extracted.  At the
          root, every atom in the digraph $G$ is in the
          checklist. The time complexity of the extractability
          tests performed at the root is thus
          $\mathcal{O}(\Sigma_{i}s_i)$.  An atom is added into the
          checklist only after the removal of some local variable
          linked to the atom. Therefore, an atom may be tested for
          its extractability for as many times as one plus the
          number of atoms with which the atom share a local
          variable. However, in the worst case smaller atoms are
          extracted before larger atoms. Thus, the $i^{th}$ atom
          can only be tested for $w_{i}+1$ times. Therefore, the
          time complexity of one derivation is
          $\mathcal{O}(\Sigma_{i}(w_{i}+1)\times s_i)$.}

Now consider (2). Since each instance of atom which is extracted at
node $nd$ is tested at most $w_{nd}+1$ times and each test costs
$s_{nd}$ unit of time. Thus, the total cost of tests in all
$\leadsto_{sqvt}$ derivations from $G$ is \(\mathcal{O}(\Sigma_{nd\in
  Fr(SPT_G)} (w_{nd}+1)*s_{nd})\).
\end{theorem}

\section{Related Work} \label{sec:related}
Apart from Cleary's original work~\cite{Cleary97}, most related works
are those on constructive negation.  The basic idea of Chan's
constructive negation approach~\cite{Chan88,Chan89} is that answers to
$\neg~Q$ are obtained by negating answers to $Q$. Given $\neg~Q$, a
frontier of a derivation tree for $Q$ is first obtained. Answers to
$\neg~Q$ are then obtained from the frontier as first-order formulae
which are interpreted in Clark's equality theory (CET). Chan's method
was formulated for logic programs in the Herbrand universe and
involves introducing disequality constraints over the Herbrand
universe. An answer to a goal by Chan's operational semantics SLD-CNF
is a set of equality and disequality constraints.  Originally, Chan's
method applied only to negative goals with finite sub-derivation trees
and worked by negating answers to the negated
sub-goal~\cite{Chan88}. Chan later extended his method by negating a
frontier of a derivation tree for the negated
sub-goal~\cite{Chan89}. The simplification procedure in Chan's method
relies on the following property of the Herbrand universe.
\[\neg\exists\yvec\zvec.(x=s(\yvec)\land Q(\yvec\zvec)) \lequiv
\forall\yvec.(x\neq s(\yvec)) \lor
\exists\yvec.(x=s(\yvec)\land\neg\exists\zvec. Q(\yvec\zvec) )\] where
$x$ is a free variable and $\yvec$ and $\zvec$ are disjoint.
Mu{\~n}oz-Hern{\'a}ndez et.\ al. refined Chan's method and
incorporated it into Ciao Prolog~\cite{Munoz-HernandezM04}. They also
implemented other negation methods~\cite{MarinoMM08} and use static
analysis to select the appropriate negation method for a negative
goal~\cite{Munoz-HernandezMH01}.

     Ma{\l}uszy\'{n}ski and N\"{a}slund put forward another approach to
     constructive negation which allows a negative goal to directly return
     fail substitutions, as its answers~\cite{Maluszynski89:NACLP}. Since
     answers to negative goals cannot in general be represented by a finite
     number of substitutions, Ma{\l}uszy\'{n}ski and N\"{a}slund's approach
     sometimes need to return an infinite number of fail substitutions.

     Drabent defines SLDFA resolution over the Herbrand
     universe~\cite{Drabent95}.  Chan's first method works only when the
     negated sub-goal has a finite number of answers. SLDFA overcomes this
     by constructing answers for the negative goal from a finite number of
     answers to the negated sub-goal.

     Fages proposes a simple concurrent pruning mechanism over standard SLD
     derivation trees for constructive negation in constraint logic
     programs~\cite{Fages97}. Two derivation trees are concurrently
     constructed. The computed answers from one of the trees are used to
     prune the nodes of the other.  Fages' method admits an efficient
     implementation as it is not necessary to deal with complex goals with
     explicit quantifiers outside the constraint part.

     Stuckey provides a constructive negation method for constraint
     logic programs over arbitrary
     structures~\cite{stuckey95}. Stuckey's method which is sound and
     complete with respect to the three-valued consequences of the
     completion of the program can be thought of as a generalisation
     of Chan's. Stuckey uses the following property of logic formulae
     in his simplification procedure.  \[ \neg\exists\yvec.(c\land
     Q)\lequiv \neg\exists\yvec.c\lor \neg\exists\yvec.(c\land Q) \]
     where $c$ is a constraint and $Q$ is a conjunction of goals. The
     method need to do a satisfiability test when combining
     $\neg\exists\yvec.c$ with other constraints.  A sufficient
     condition for applying Stuckey's method is that the constraint
     domain has the admissible closure property, i.e.,
     $\neg\exists\yvec.c$ for any admissible constraint $c$ can be
     rewritten as a disjunction of admissible
     constraints~\cite{stuckey95}. Dovier et.\ al. prove that the
     admissible closure property is also a necessary condition for an
     effective implemention of the method~\cite{DovierPR00}.

\comments {The extraction method presented in this paper may be used
  as part of a constructive negation system. The method is correct in
  that its rewrite rules preserve logical equivalence.  The method is
  also complete in the sense that it doesn't throw away any answer to
  a negative goal; it simply suspends when none of its rewrite rules
  is applicable. The current implementation includes \existence
  properties for arithmetic constraints. It also allows the programmer
  to provide existence properties for other predicates.  There has
  been much research into constructive
  negation~\cite{ICLP94:Bossi,ICLP94:Bruscoli,Paola94,Chan88,Chan89,Drabent95,Fages97,Maluszynski89:NACLP,Marchiori96,Moreno-Navarro94,
    Moreno-Navarro96,Moreno-Navarro:2000,Munoz-HernandezMM04,RamirezF93,stuckey95}.}
Constructive intensional negation was studied
in~\cite{ICLP94:Bossi,Paola94,ICLP94:Bruscoli,MarinoMM08,Munoz-HernandezMM04}.
Marchiori~\cite{Marchiori96} addresses the termination of logic
programs with respect to constructive negation.  Lobo~\cite{Lobo90}
studies constructive negation for disjunctive logic programs.
Ram\'{\i}rez and Falaschi~\cite{RamirezF93} and
Moreno-Navaro~\cite{Moreno-Navarro94,Moreno-Navarro96,Moreno-Navarro:2000}
extend constructive negation for functional logic programs. Dovier
et.\ al. extends Chan's method to $CLP(SET)$ where $SET$ is the domain
of hereditarily finite sets~\cite{DovierPR01}.  $SET$ does not satisfy
the admissible closure property and hence the constructive negation
method is complete only for a subset of $CLP(SET)$~\cite{DovierPR00}.

We now compare our method with Chan's and Stuckey's using
Ex.~\ref{ex:append}. QVT rewrites \(\neg\exists
z\!\!:\!\!\tlist(\beta).(append(x\!\!:\!\!\tlist(\beta),y\!\!:\!\!\tlist(\beta),z),p(z))
\) to \(
append(x\!\!:\!\!\tlist(\beta),y\!\!:\!\!\tlist(\beta),z\!\!:\!\!\tlist(\beta)),\neg~p(z)
\).  Both Chan's method and Stuckey's first construct an SLD
derivation tree of $append(x,y,z),p(z)$ and collect a frontier of the
SLD derivation, say,
\[\left\{\begin{array}{c} (x=[],y=z,p(z)),\\ (x=[h|x'],y=y',z=[h|z'],append(x',y',z'),p(z))\end{array}\right\}\]
Then the negation of this frontier is simplified and put into its
disjunctive normal form. This gives rise to the following four conjunctive
formulae.
\begin{itemize}
\item [(1)] \(x\neq[],\forall h,x'.(x\neq[h|x'])\)
\item [(2)] \(x\neq[],x=[h|x'],\neg\exists z'.(append(x',y,z'),p([h|z']))\)
\item [(3)] \(x=[],\forall h,x'.(x\neq[h|x']),\neg p(y)\)
\item [(4)] \(x=[],x=[h|x'],\neg p(y),\neg \exists z'.(append(x',y,z'),p([h|z'])) \)
\end{itemize}
Stuckey's method derives (2) and (3) because the constraint parts of
(1) and (4) are unsatisfiable.  Chan's method derives (1),(2) and (3)
as it only tests satisfiability of atomic constraints. The constraint
part of (4) is failed by unification in Chan's method as \([]\) is not
unifiable with \([h|x']\). Neither of these methods is effective as
(2) is as complex as the original goal.  The \eu property allows us to
obtain a simpler derived goal without making use of SLD derivation,
and to eliminate unsatisfiable derived goals without satisfiability
tests. Similar comparison can be made between our's and methods
in~\cite{Drabent95,Fages97,Maluszynski89:NACLP} since they all
construct a frontier of an SLD derivation tree for
$append(x,y,z),p(z)$.

\section{Conclusion} \label{sec:conc}

We have presented a simplification method that uses typed \existence
properties to rewrite negative goals. The method strictly generalizes
an earlier work that uses functional dependencies to rewrite negative
goals. A typed existence property generalizes a functional dependency
in that the domains of both input and output parameters can be
restricted to sub-domains and moreover one input value may correspond
to more than one output values. The method consists of rewrite rules
one for each kind of typed \existence properties. The rewrite rules
doesn't involve an SLD-derivation of the negated sub-goal nor an
explicit satisfiability test.

We have described an implementation of the method and analyzed its
complexity. The implementation uses a digraph and a worklist to
represent a negative goal so as to avoid futile extractability tests
of atoms in the negative goal.  An algorithm is presented that does
the extractability test given an atom and an \existence property and
introduces new local variables into the atom to make it satisfy the
\existence property. The complexity of the algorithm is linear in the
size of the atom.

\subsection*{Acknowledgement} 
We would like to thank anonymous referees for their constructive
comments and suggestions.


\begin{thebibliography}{10}

\bibitem{Aggoun+95}
A.~Aggoun et.\ al.\
\newblock {\em $ECL^{i}PS^{e}$ 3.5 User Manual}.
\newblock ECRC Munich, Germany, December 1995.

\bibitem{ICLP94:Bossi}
A.~Bossi, M.~Fabris, and M.C. Meo.
\newblock A bottom-up semantics for constructive negation.
\newblock In \cite{ICLP94}, pages 520--534.

\bibitem{Brisset+94}
P.~Brisset et.\ al.\
\newblock {\em $ECL^{i}PS^{e}$ 3.4 Extensions User Manual}.
\newblock ECRC Munich, Germany, July 1994.

\bibitem{ICLP94:Bruscoli}
P.~Bruscoli, A.~Dovier, E.~Pontelli, and G.~Rossi.
\newblock Compiling intensional sets in {CLP}.
\newblock In \cite{ICLP94}, pages 647--661.



\bibitem{Paola94}
P.~Bruscoli, F.~Levi, G.~Levi, and M.C. Meo.
\newblock Compilative constructive negation in constraint logic programs.
\newblock {\em Lecture Notes in Computer Science}, 787:52--67, 1994.

\bibitem{ICLP94}
M.~Bruynooghe, editor.
\newblock {\em Proceedings of the Eleventh International Conference on Logic
  Programming}. The MIT Press, 1994.

\bibitem{Chan88}
D.~Chan.
\newblock Constructive negation based on the completed database.
\newblock In  \cite{ICLP88}, pages 111--125.

\bibitem{Chan89}
D.~Chan.
\newblock An {E}xtension of {C}onstructive {N}egation and its {A}pplication in
  {C}oroutining.
\newblock In  {\em Proceedings of
  the 1989 North American Conference on Logic Programming}, pages 477--496. The
  MIT Press, 1989.

\bibitem{Cleary97}
J.G. Cleary.
\newblock Constructive negation of arithmetic constraints using data-flow
  graphs.
\newblock {\em Constraints}, 2:131--162, 1997.

\bibitem{ClearyLu98}
J.G. Cleary and L.~Lu.
\newblock Constructive negation using typed existence properties.
\newblock  {\em Lecture Notes in Computer  Science}, 1490:411--426, 1998.

\bibitem{DovierPR00}
A. Dovier and E. Pontelli and G. Rossi.
\newblock A necessary condition for Constructive Negation in Constraint
               Logic Programming.
\newblock {\em Inf. Process. Lett.}, 74 (3\&4):147-156, 2000.


\bibitem{DovierPR01}
A.~Dovier, E.~Pontelli and G.~Rossi.
\newblock Constructive Negation and Constraint Logic Programming with
               Sets.
\newblock {\em New Generation Comput.}, 19 (3):209-256, 2001.

\bibitem{Drabent95}
W.~Drabent.
\newblock {What is failure? An approach to constructive negation}.
\newblock {\em Acta Informatica}, 32:27--59, 1995.

\bibitem{Fages97}
F.~Fages.
\newblock Constructive negation by pruning.
\newblock {\em Journal of Logic Programming}, 32(2):85--118, 1997.


\bibitem{FruhwirthSVY:LICS91}
T. Fr\"{u}hwirth, E. Shapiro, M.Y. Vardi and E. Yardeni.
\newblock Logic Programs as Types for Logic Programs.
\newblock In {\em Proceedings of the Sixth Annual IEEE
                  Symposium on Logic in Computer Science},       pages 300-309,
The IEEE Computer Society Press,       1991.



\bibitem{ICLP88}
R.~A. Kowalski and K.~A. Bowen, editors.
\newblock {\em Proceedings of the Fifth International Conference and Symposium
  on Logic Programming}. The MIT Press, 1988.

\bibitem{Lobo90}
Jorge Lobo.
\newblock On constructive negation for disjunctive logic programs.
\newblock In  {\em Proceedings
  of the 1990 North American Conference on Logic Programming}, pages 704--718,
 The MIT Press, 1990.

\bibitem{Maluszynski89:NACLP}
J.~Ma{\l}uszy{\'n}ski and T.~N{\"a}slund.
\newblock Fail {S}ubstitutions for {N}egation as {F}ailure.
\newblock In  {\em Proceedings of
  the 1989 North American Conference on Logic Programming}, pages 461--476. The MIT
  Press, 1989.

\bibitem{Marchiori96}
E. Marchiori.
\newblock On termination of general logic programs w.r.t. constructive
  negation.
\newblock {\em Journal of Logic Programming}, 26(1):69--89, 1996.

\bibitem{Mycroft:OKeefe:84}
A. Mycroft and R.A. O'Keefe.
\newblock A Polymorphic Type System for Prolog.
\newblock {\em Artificial Intelligence}, 23(3): 295--307, 1984.

\bibitem{MarinoMM08}
J.~Mari{\~n}o and  J.~J. Moreno-Navarro and
               S. Mu{\~n}oz-Hern{\'a}ndez.
\newblock Implementing Constructive Intensional Negation.
\newblock {\em New Generation Comput.}, 27 (1):25-56, 2008.

\bibitem{Moreno-Navarro94}
J.~J. Moreno-Navaro.
\newblock Default rules: An extension of constructive negation for
  narrowing-based languages.
\newblock In \cite{ICLP94}, pages 535--549.

\bibitem{Moreno-Navarro96}
J.~J. Moreno-Navarro.
\newblock Extending constructive negation for partial functions in lazy
  functional-logic languages.
\newblock {\em Lecture Notes in Artificial Intelligence}, 1050:213--228, 1996.

\bibitem{Moreno-Navarro:2000}
J.~J. Moreno-Navarro and S.~Mu{\~n}oz-Hern{\'a}ndez.
\newblock How to incorporate negation in a {Prolog} compiler.
\newblock {\em Lecture Notes in Computer Science}, 1753:124--139, 2000.

\bibitem{Munoz-HernandezM04}
S.~Mu{\~n}oz-Hern{\'a}ndez and J.J. Moreno-Navarro.
\newblock Implementation Results in Classical Constructive Negation.
\newblock \newblock {\em Lecture Notes in Computer Science}, 3132:284-298, 2004.


\bibitem{Munoz-HernandezMH01}
S.~Mu{\~n}oz-Hern{\'a}ndez, J.~J. Moreno-Navarro and  M.~V. Hermenegildo.
\newblock Efficient Negation Using Abstract Interpretation.
\newblock {\em Lecture Notes in Computer Science}, 2250:485-494, 2001.

\bibitem{Munoz-HernandezMM04}
S.~Mu{\~n}oz-Hern{\'a}ndez, J.~Mari{\~n}o, and J.J. Moreno-Navarro.
\newblock Constructive intensional negation.
\newblock {\em Lecture Notes in Computer Science}, 2998:39--54, 2004.



\bibitem{RamirezF93}
M.~J. Ram\'{\i}rez and M.~Falaschi.
\newblock Conditional {N}arrowing with {C}onstructive {N}egation.
\newblock {\em Lecture Notes in Artificial Intelligence}, 660:59--79, 1993.

\bibitem{stuckey95}
P.J. Stuckey.
\newblock Negation and constraint logic programming.
\newblock {\em Information and Computation}, 118:12--33, 1995.


\bibitem{Yardeni:Shapiro:91}
E. Yardeni and E. Shapiro.
\newblock A Type System for Logic Programs.
\newblock {\em Journal of Logic Programming}, 10(2): 125--153, 1991.



\end{thebibliography}
\end{document}